\documentclass[aps,
               prb,
               twocolumn,
               10pt,
               longbibliography,
               floatfix,
               nofootinbib,
               superscriptaddress]{revtex4-2}
\usepackage{cancel}
\usepackage{physics}
\usepackage[english]{babel}
\usepackage[utf8]{inputenc}
\usepackage{amssymb}
\usepackage{amsmath}
\usepackage{multirow}
\usepackage{float}
\usepackage[pdftex]{graphicx}
\usepackage{xspace}
\usepackage{dsfont}
\usepackage{soul}
\usepackage{bm}
\usepackage[normalem]{ulem}
\usepackage{hyperref}
\hypersetup{
    colorlinks,%
    citecolor=blue,%
    filecolor=blue,%
    linkcolor=blue,%
    urlcolor=blue
}
\usepackage[usenames, dvipsnames]{color}

\graphicspath{{.}{Figuras/}} 

\begin{document}


\title{Approaching the Quantum Speed Limit in Quantum Gates with Geometric Control
}
\author{François Impens}
\affiliation{Instituto de F\'{\i}sica, Universidade Federal 
Rio de Janeiro, 21941-972 Rio de Janeiro, RJ, Brazil}

\author{David Guéry-Odelin}
\affiliation{Laboratoire Collisions Agrégats Réactivité, UMR 5589, FERMI, Université de Toulouse, CNRS, 118 Route de Narbonne, 31062 Toulouse CEDEX 09, France.}

\date{July 2025}

\begin{abstract}
We present a geometric optimization method for implementing quantum gates by optimally controlling the Hamiltonian parameters, with the goal of approaching the Mandelstam-Tamm Quantum Speed Limit (MT-QSL). Achieving this bound requires satisfying precise geometric conditions that govern the evolution of quantum states. We extend this geometric framework to quantum unitary operators in arbitrary dimensions and analyze the conditions necessary for saturation of the bound. Additionally, we show that the quantum brachistochrone, when generalized to operators, does not, in general, saturate the MT-QSL bound. Finally, we propose a systematic optimal control strategy based on geometric principles to approach the quantum speed limit for unitary driving. We illustrate this optimization method on a set of four well-known two-qubit quantum gates. Our procedure significantly reduces the deviation from the optimal quantum speed limit while preserving high quantum fidelity.
\end{abstract}

\maketitle

    Approaching optimality in quantum driving is both a challenging and crucial task for quantum information processing. Reducing the evolution time while keeping resources constant enhances efficiency and can increase the resilience of quantum protocols to external perturbations. The concept of Quantum Speed Limit (QSL) provides a quantitative bound for this requirement. The QSL literature commonly presents two main criteria, depending on the specific problem: a global criterion that defines the minimal time for an initial state to reach an orthogonal subspace, and a local criterion that bounds the instantaneous angular separation between the initial and evolving states. This concept has been a subject of growing interest, leading to numerous extensions~\cite{QSL45,QSLEberly73,Vaidmann92,QSLVittorio03,QSL05,QSLLevitin09,QSL12,QSLTight18,PhysRevLett.129.140403}, with the recent generalization to open ~\cite{QSLRuynet13,QSLDelCampo13,QSLImpens21,QSLNonUnitary22}, non-Markovian~\cite{QSLDeffner13} and non-linear~\cite{QSLDeffner22,QuantSynchro23} quantum systems. Among recent developments, we highlight efforts to establish unified geometric definitions of the QSL~\cite{QSLGeometricPRX16}, the investigation of the relation of the QSL with the system quantumness~\cite{QSLQuantClass18,QSLNonQuantum18}, with interference effects~\cite{QSLSuperRadiance20,QSLCoh24}, or with statistical properties~\cite{QSLStat18,Q5LPRE21} of the system, and finally an effort to approach the QSL in concrete experimental setups~\cite{Ness21,QSLNatCom24}. The QSL is fundamentally linked to the geometry of quantum state space~\cite{QSLAnandan90} and involves identifying quantum trajectories that correspond to geodesics in the Fubini-Study metric. When formulated as a quantum brachistochrone problem, it allows for analytical~\cite{Carlini06,Carlini07} or approximate~\cite{Brachistochrone15} solutions for specific classes of Hamiltonians. However, optimal solutions to the quantum brachistochrone problem do not always saturate the QSL bounds associated with the evolution of a single quantum state~\cite{QSLBrachistochrone23}. While a significant body of literature has focused on QSLs for the evolution of individual states, the exploration of QSLs governing the evolution of sets of quantum states remains in its early stages~\cite{QSLUnitary12,QSLUnitaryDelCampo22,LuisGarciaPRX22,DelCampoQSL23,QSLUnitary24,QSLUnitary24b}.

The goal of this paper is to present an alternative derivation of such bounds, inspired by the approach of L. Vaidman~\cite{Vaidmann92}, which also introduces a geometric optimization method for implementing time-optimal unitary quantum gates.
More specifically, we identify two necessary criteria for optimal driving: one involves a phase condition, while the other has a geometric interpretation. Indeed, the deviation from the optimal Mandelstam–Tamm QSL bound grows as the evolution operator departs from a specific time-dependent plane in the Hilbert space of quantum operators.

In Sec.~\ref{Sec:QSLbounds}, we generalize the Mandelstam-Tamm bound on the quantum speed limit using an approach similar to that developed by L. Vaidman \cite{Vaidmann92}. We then discuss the generalization of the Margolus-Levitin bound for time evolution operators. In Sec.~\ref{SecSaturation}, we explore the conditions under which the bound is saturated, illustrating these with various examples, and analyze the quantum brachistochrone from the QSL perspective. In Sec.~\ref{SecOCT}, we detail the quantum optimal driving procedure designed to achieve the maximal Quantum Speed Limit, based on Pontryagin's maximum principle and using geometrically derived cost functionals. Finally, in Sec.~\ref{Secexamples}, we illustrate our protocol for various unitary gates in a four-dimensional space. We show the importance of the control over a large number of Hamiltonian parameters, compare the efficiency of the algorithm with respect to different cost functionals, and discuss the results of another optimization scheme.

\section{Quantum Speed Limits for unitary operators}
\label{Sec:QSLbounds}

In this paragraph, we derive the two standard bounds for the quantum speed limit in the context of quantum control of time evolution operators.
We consider a $d$-dimensional Hilbert space $\mathcal{H}$ and its associated space of linear operators $\mathcal{L}(\mathcal{H})$, on which we define the scalar product  $\langle U_1, U_2 \rangle = \frac 1 d {\rm Tr} \left[ U_1^{\dagger} U_2 \right] $. Unitary operators $U$ have unit norm $\langle U, U \rangle =1$. This metric provides a suitable criterion for implementing a target quantum gate since for any unitary operators $U$ and $U_{\rm target}$, $\langle U_{\rm target}, U \rangle~=~1~\Longrightarrow~U=U_{\rm target}$.

\subsection{Mandelstamm and Tamm bound for quantum operators}

Using geometric arguments, L. Vaidman derived the minimal time required to evolve a given initial quantum state into an orthogonal one ~\cite{Vaidmann92}.  Following a similar approach, we derive a QSL for the full quantum evolution operator and establish a lower bound on the evolution time required to reach a target quantum gate $U_{\rm target}$.

Given an arbitrary, possibly time-dependent, Hamiltonian $H(t)$ and a unitary evolution operator $U(t)$, we decompose $H(t)U(t)$ along the direction spanned by $U(t)$ and its orthogonal subspace as follows:
 \begin{equation}
\label{eq:decomposition}
H(t) U(t) = \lambda_{/\!/}(t) U(t) + \lambda_{\perp}(t) U_{\perp}(t)
\end{equation}
where $U_{\perp}(t)$ is a linear operator such that $\langle U(t), U_{\perp}(t) \rangle=0$ and  $\langle U_{\perp}(t),U_{\perp}(t) \rangle=1$ (note that $U_{\perp}(t)$ is in general non-unitary). Projecting both sides of the equation onto the operator $U(t)$, we obtain $\lambda_{/\!/}(t)= \frac 1 d {\rm Tr} [ U^{\dagger}(t) H(t) U(t)] = \frac {1} {d}  {\rm Tr} [H(t)]  =\langle H(t) \rangle$ - where we define $\langle O \rangle = \frac {1} {d}  {\rm Tr} [O ] $ for any operator $O$.  Next, we take the scalar product with the Hermitian conjugate $H(t) U(t)$ on both sides of Eq.~\eqref{eq:decomposition}, and readily obtain $\frac 1 d {\rm Tr} [ H(t)^2 ] = |\lambda_{/\!/}(t)|^2 + |\lambda_{\perp}(t)|^2$ where we have used $H(t)^\dagger=H(t)$, $\langle U(t),U(t) \rangle=\langle U_\perp(t),U_\perp(t)  \rangle=1$ and $\langle U_{\perp}(t), U(t) \rangle= \langle U(t), U_\perp(t) \rangle = 0.$ As a result, we find $|\lambda_{\perp}(t)|^2= \langle H(t)^2 \rangle - \langle H(t) \rangle^2$, i.e. $|\lambda_{\perp}(t)|=\Delta H(t).$ Consequently, the operator $U_{\perp} $ can be expressed as
\begin{equation}
\label{eq:Uperpdefinition}
 U_{\perp}(t) = \frac {1} {\Delta H(t)} \left( H(t) U(t) - \langle H(t) \rangle U(t) \right)
\end{equation}
as long as $\Delta H(t) \neq 0$. 

Using the decomposition above, the Schr\"odinger equation for the evolution of $U(t)$ reads:
\begin{equation}
\dot{U}= - \frac {i} {\hbar} \left( \langle H \rangle \: U + \Delta H \: U_{\perp} \right).
\end{equation}
 The initial evolution operator is the identity matrix $U_0=\mathbb{I}$. The time derivative of its overlap with the evolved quantum operator can be estimated as:
\begin{eqnarray}
\label{eq:timederivativesquare}
 \frac {d | \langle  U_0, U(t) \rangle |^2} {dt} = \frac {- 2 \Delta H(t)} {\hbar} {\rm Re} \left[  i  \langle U(t),  U_0 \rangle  \langle  U_0, U_{\perp}(t) \rangle \right] \nonumber \\
\end{eqnarray}

We have omitted the term proportional to the average energy, which corresponds to a purely imaginary number.  The initial evolution operator $U_{0}$ can be decomposed as
\begin{equation}
\label{eq:decomposition2}
 U_0 = \langle U(t), U_{0} \rangle U(t)+ \langle U_{\perp}(t),  U_{0} \rangle U_{\perp}(t)+  U_{\perp \perp}(t)
\end{equation}
where $U_{\perp \perp}(t)$ is a linear operator that is orthogonal to both $U(t)$ and $U_{\perp}(t)$. We obtain 

\begin{eqnarray}
\label{eq:inequality}
 |\langle  U_0, U_{\perp}(t) \rangle| & = & \sqrt{1- | \langle  U_0, U(t) \rangle|^2- ||U_{\perp \perp}(t)||^2} \nonumber \\  & \leq &   \sqrt{1- | \langle U_{0}, U(t) \rangle|^2}.
\end{eqnarray}
where the inequality is saturated if and only if 
\begin{equation}
\label{eq:planecondition0}
||U_{\perp \perp}(t)||^2=0
\end{equation}
i.e.~if $ U_0$ belongs to the plane spanned by $\{ U(t),H(t)U(t) \}$ at all times.

Combining Eq.~\eqref{eq:timederivativesquare} with the inequality~\eqref{eq:inequality}, we obtain:
\begin{equation}
 \left| \frac {d | \langle  U_0, U(t) \rangle |^2} {dt} \right| \leq \frac {2 \Delta H(t)} {\hbar} | \langle  U_0, U(t) \rangle| \sqrt{1- | \langle  U_0, U(t) \rangle|^2}.
\label{inequalityqsl}
\end{equation}
By setting $\cos \phi= | \langle  U_0, U(t) \rangle |$, the inequality~\eqref{inequalityqsl} can be expressed as an upper bound on the time derivative of the angle $\phi$:
\begin{equation}
\label{eq:QSL}
|\dot{\phi}| \leq \frac { \Delta H} {\hbar}.
\end{equation}
By construction, the initial angle is $\phi(0)=0$. For a time-independent Hamiltonian $H$, the minimal time $\tau$ to reach a given target quantum gate $U_{\rm target}$ is given by:
\begin{equation}
    \tau = \frac {\hbar \phi_{\rm target}} {\Delta H} \quad \text{where} \quad \phi_{\rm target} = {\rm Arccos}\left[ \frac {|{\rm Tr}[U_{\rm target}]|} {d} \right].
\end{equation}

For a target gate with vanishing trace, $\phi_{\rm target}=\pi/2$. In this specific case, we retrieve a result analogous to the Mandelstamm-Tamm time bound for reaching an orthogonal quantum state.

Saturation of this QSL requires that two conditions be simultaneously fulfilled
\begin{eqnarray}
\label{eq:twoQSLconditions}
| \langle U_0, U(t) \rangle|^2+|\langle  U_0, U_{\perp}(t) \rangle|^2 & = & 1 \nonumber \\
{\rm Re} \left[  \langle  U(t),   U_0 \rangle  \langle U_0, U_{\perp}(t) \rangle  \right]  & = &  0 
\end{eqnarray}
The first condition states that $U_{\perp \perp}(t)=0,$ meaning that the time derivative $\dot{U}(t)$ must always lie in the plane spanned by $\{  U_0, U(t) \}$. Since the evolution operator $U(t)$ must remain unitary at all times, this imposes strong constraints on the possible quantum paths saturating the QSL bound. In practice, it is not guaranteed that such a path exists in Hilbert spaces with dimensions larger than two. 

In such cases, one can quantify the deviation from the QSL when these conditions are not satisfied. Using
$||U_{\perp \perp}(t) ||=\sqrt{\sin^2 \phi(t)- |\langle  U_0,U_{\perp}(t) \rangle|^2}$ and $\beta(t)= {\rm Arg}[\langle U(t), U_0 \rangle \langle U_0, U_{\perp}(t) \rangle]$, one has
\begin{equation}
\label{eq:imperfectvelocity}
 |\dot{\phi}(t)| = \frac {\Delta H(t)} {\hbar} \sqrt{ 1 - \frac { ||U_{\perp \perp}(t) ||^2} {\sin \phi(t)^2}} |\sin \beta(t)|.
\end{equation}
The ideal case of QSL saturation corresponds to $||U_{\perp \perp}(t) ||=0$ (plane condition) and to $\beta(t)=\pm \frac {\pi} {2}$~(phase condition) at all times. An optimization method based on these two relations will be presented in Sec.~\ref{SecOCT}.

\subsection{Margolus-Levitin quantum speed limit}

the Margolus–Levitin bound ~\cite{Margolus98} can also be recovered within this operatorial formalism. To this end, we consider a time-independent Hamiltonian $H_0= \sum_{n=1}^{d} E_n | n \rangle \langle n |$ with $E_1 \leq E_2 \leq... \leq E_d$ possibly with degenerate eigenvalues. The associated evolution operator reads $U(t)= \sum_{n=1}^{d} e^{- \frac {i} {\hbar} E_n t} | n \rangle \langle n|,$ and one finds
\begin{equation}
D(t)= \frac 1 d {\rm Tr} \left[ U_0^{\dagger} U(t) \right]= \frac {1} {d}  \sum_{n=1}^{d} e^{- \frac {i} {\hbar} E_n t},
\end{equation}
where we take the initial evolution operator as $U_0= \mathbb{I}$ the same reasoning applies to any other gate $U_f$). One can proceed in the same way as in the proof for single states (with here $|c_n|^2 \equiv 1/d$). Thus, we have that
\begin{eqnarray}
\label{eq:sum_energies}
{\rm Re} \left[ D(t) \right] & = & \frac {1} {d}  \sum_{n=1}^{d} \cos \left( \frac {E_n} {\hbar} t \right)  \\
& \geq &  \frac {1} {d} \sum_n \left[ 1- \frac {2} {\pi} \left( \frac {E_n t} {\hbar} + \sin \left( \frac {E_n t} {\hbar} \right) \right) \right],  \nonumber 
\end{eqnarray}
where we have used that 
\begin{equation}
\label{eq:inequality2}
\cos(x) \geq 1-\frac {2} {\pi} \left( x + \sin(x) \right)
\end{equation}
valid for $x \geq 0$, which becomes an equality if and only if $x=0,\pi$. Finally, the right-hand side of the inequality~\eqref{eq:sum_energies} can be expressed in terms of the average energy and imaginary part of $D(t)$, so that:
\begin{eqnarray}
{\rm Re} \left[ D(t) \right] \geq 1 - \frac {2} {\pi} \langle H_0 \rangle \frac {t} {\hbar} +  \frac {2} {\pi} {\rm Im}[D(t)].
\end{eqnarray}
This can be interpreted in terms of minimum time $\tau$ to reach a given target quantum gate $U_{\rm target}$:
\begin{equation}
\label{eq:targetquantumgate}
\tau \geq  \frac {\hbar \pi} {2  \langle H_0 \rangle} \left( 1 - D^R_{\rm target} + \frac {2} {\pi} D^I_{\rm target}  \right),
\end{equation}
where we have noted $D^R_{\rm target} = \frac 1 d {\rm Re}\left[ {\rm Tr}[U_{\rm target}] \right]$ and $D^I_{\rm target} = \frac 1 d {\rm Im}\left[ {\rm Tr}[U_{\rm target}] \right] $. For the particular case where the target quantum gate has a null trace, one obtains  
\begin{equation}
\label{eq:QSLMLbound}
\tau \geq \frac {  \pi\hbar} {2\langle H_0 \rangle},
\end{equation}
which is the quantum-operator equivalent of the Margolus-Levitin bound for individual quantum states.

 When  ${\rm Tr}[U_{\rm target}]= 0$, a tighter bound can be derived. The argument is valid for any constant and positive Hamiltonian $H_0$. The replacement $H_0' = H_0 - E'$ merely adds a global phase factor and thus leaves invariant the time $\tau$  such that $D(\tau)=0$. As long as $E' \leq E_1$ to guarantee the positivity of the Hamiltonian, the argument above yields a similar bound with $\langle H_0' \rangle = \langle H_0\rangle - E'$ instead of $\langle H_0 \rangle.$ The tighter bound is obtained with $E'=E_1,$ so that one can replace $\langle H_0 \rangle$ by $\langle H_0 \rangle-E_1$ in the denominator of Eq.~\eqref{eq:QSLMLbound}. When ${\rm Tr}[U_{\rm target}] \neq 0$, adding a constant energy term can alter the time $\tau$ required to achieve the quantum gate. Therefore, the denominator in Eq.~\eqref{eq:targetquantumgate} must remain $\langle H_0 \rangle$.

To saturate the Margolus-Levitin bound, the only admissible values of $x=E_n \tau$ in inequality \eqref{eq:sum_energies} are $\{0,\pi\}$. Consequently, only quantum gates with a real-valued trace can be attained through a path that saturates the Margolus–Levitin bound.

 Let's reconsider the case of a target gate where $D(\tau)=\frac {1} {d} {\rm Tr}[U_{\rm target}]=0$. One might wonder under which conditions the QSL bound is attained. To simplify the discussion, we consider a Hamiltonian $H_0'=H_0-E_1$  with a ground state of zero energy, and the associated quantity $D(t)$. All the terms in the sum of Eq.~\eqref{eq:sum_energies} must saturate the inequality Eq.~\eqref{eq:inequality2}. The first term, by construction, corresponds to $x=0$, and there can be at most one additional term corresponding to $x=(E_{j}-E_1) \tau /\hbar=\pi$ for some $j > 1$. The only way to saturate the bound is therefore to choose the evolution time as $\tau= \pi/\hbar(E_j-E_1),$ and to have a Hamiltonian of the form:
\begin{equation}
H_0 = \sum_{n \in \mathcal{S}_1} E_1 | n \rangle \langle n | +\sum_{n \in \mathcal{S}_j} E_j | n \rangle \langle n |,
\end{equation}
where $\{ \mathcal{S}_1, \mathcal{S}_j \}$ is a partition of the eigenstate basis in two degenerate subspaces of energies $\{E_1,E_j \}$. 

The coincidence of the MT and ML QSL bounds occurs when $\Delta H=\langle H \rangle - E_1 $. Since $\langle H \rangle - E_1 = \frac n d (E_j-E_1)$ and  $\Delta H = \frac 1 d \sqrt{n(d-n)} (E_j-E_1)$ (with $n={\rm card} \mathcal{S}_j$), this reduces to the simple condition $n=d/2$. In this case, the two bounds coincide and are saturated when the quantum system exhibits dynamics essentially identical to those of a two-level system.

\section{Saturation of the MT-QSL bound for specific families of unitary dynamics}
\label{SecSaturation}
To illustrate the constraints imposed by the Mandelstam-Tamm QSL for unitary operators, we investigate the potential saturation of this quantum speed limit in low-dimensional systems and discuss the brachistochrone approach within the QSL framework.

\subsection{Saturation of the MT-QSL in a 2D system}

In a two-dimensional system, a generic quantum state is fully defined by two angles (up to an arbitrary phase), in agreement with the conventional Bloch sphere representation. As a result, when defining a general unitary operator, the mapping of the first state of the canonical basis corresponds to two degrees of freedom. The second basis state is necessarily mapped to an orthogonal state, but its phase can still be chosen independently. Therefore, quantum gates in a two-dimensional Hilbert space are parametrized by three degrees of freedom.

Here, we examine the saturation of the local QSL criterion~\eqref{eq:QSL} within a family of unitary dynamics parameterized by two control parameters $U(\theta(t), \varphi(t)) = e^{ - \frac i 2 \varphi(t)} | \psi_+(\theta(t)) \rangle \langle 0 |+e^{ \frac i 2 \varphi(t)}| \psi_-(\theta(t)) \rangle \langle 1 |$ with the two orthogonal states  $| \psi_+(\theta) \rangle
= \cos (\theta / 2) |0 \rangle + \sin (\theta/ 2) |1 \rangle$ and $| \psi_-(\theta) \rangle =  -\sin (\theta/ 2) |0 \rangle + \cos (\theta/ 2) |1 \rangle$. This corresponds to a subset of the possible unitary gates in a  two-dimensional system, but the extension to the more general case, which includes all possible unitary operators, is straightforward. We now determine the dynamics of the parameters $(\theta(t), \varphi(t))$ such that the QSL~\eqref{eq:QSL} is saturated at all times. The Hamiltonian is determined by $H(t) = i \hbar  \dot{U}(t) U^{\dagger}(t)$ and is given by $H(t)= \frac \hbar 2 \left[  \dot{\theta}  U(t)  \: \sigma_y U^{\dagger}(t)  + \dot{\varphi} \: U(t) \sigma_z U^{\dagger}(t) \right]$ where $\sigma_{x,y,z}$ are the usual Pauli operators. We obtain  $\Delta H= \frac \hbar 2 \sqrt{\dot{\theta^2}+\dot{\varphi}^2}.$ On the other hand, the overlap with the initial evolution operator is given by $\langle U_0, U(t) \rangle = \cos \left( \frac 1 2 \theta(t) \right) \cos \left(  \frac 1 2 \varphi(t) \right)$. By setting $\cos \phi = \langle U_0, U(t) \rangle$, we find
\begin{equation}
\dot{\phi} = \frac {\dot{\theta}  \sqrt{1-x^2} y + \dot{\varphi} x \sqrt{1-y^2}  } {2 \sqrt{1-x^2 y^2}}
\end{equation}
with $x=\cos \left( \frac  {\theta(t)} {2} \right) $ and $y=\cos \left(  \frac {\varphi(t)} {2} \right)$ (we omit the time-dependence of $x(t)$ and $y(t)$ to simplify the notation). This can be recast as $\dot{\phi}=\mathbf{a} \cdot \mathbf{b} /(2 \sqrt{1-x^2 y^2}) $ with the column vectors $\mathbf{a}= (\dot{\theta},\dot{\varphi})^T$ and $\mathbf{b}=(\sqrt{1-x^2} y,x \sqrt{1-y^2})^T$. The Cauchy-Schwarz inequality yields 
\begin{equation}
\label{eq:inequality1}
\dot{\phi} \leq \frac {\Delta H} {\hbar} \frac {\sqrt{x^2+y^2-2 x^2 y^2}} {\sqrt{1-x^2 y^2}} \, .
\end{equation}
Finally, the inequality 
\begin{equation}
\label{eq:inequality2}
\sqrt{x^2+y^2-2 x^2 y^2} \leq \sqrt{1-x^2 y^2}
\end{equation}
holds because $|x|$ and $|y|$ are both within the range  $[0,1]$.

The QSL is attained when both the inequalities in Eqs.~ \eqref{eq:inequality1} and \eqref{eq:inequality2} are simultaneously saturated. Equation \eqref{eq:inequality2} saturates if either $|x|=1$ or $|y|=1$. If $|x|=1$ (meaning $\theta(t)=0$) and $0 \leq |y| < 1$, then the Cauchy-Schwarz inequality saturates if and only if the vectors $\mathbf{a}$ and $\mathbf{b}$ are collinear, which means $\dot{\theta}(t)=0$.
Conversely, if $|y|=1$ (meaning $\varphi(t)=0$) and $0 \leq x < 1$, saturation occurs if and only if $\dot{\varphi}(t)=0$. To sum up, the paths saturating QSL~\eqref{eq:QSL} follow the great arcs of the sphere in the parameter space $(\theta,\varphi)$. In the context of quantum operators, we find a well-known result for the evolution of individual quantum states~\cite{QuantumGeometryBook}: in two-dimensional systems, the geodesics associated to the Fubini-Study metric coincide with the great arcs of the Bloch sphere. To achieve QSL saturation, only one of the two angles $(\theta(t),\varphi(t))$ can evolve at a time. This example shows that saturating the QSL severely constrains the choice of possible quantum paths.


\subsection{Investigation of the QSL limit in 3-level systems with Rodrigues formula}

In Ref.~\cite{ReverseEngStirap17}, a method was proposed to control the dynamics of a three-level system by separating the evolution into two components: population changes, which are parameterized using Rodrigues' rotation formula~ [Eqs.(\ref{eq:RodriguesRotationFormula},\ref{eq:decompositionUnophase})], and phase changes.  This approach was used to inversely design the Hamiltonian in a three-level system with limited access to the couplings. 


Similarly, we focus here on a three-state system and the family of unitary operators ${ U_r(t) }$ that govern population transfer. Specifically, we consider mappings between quantum states with positive real-valued coefficients. We denote $\mathcal{S}=\{ |\psi \rangle | \forall n \in \{1,2,3\} \:  \langle n | \psi \rangle \in \mathbb{R}^+ \}$  as the corresponding set of quantum states. In other words, if $| \psi(0) \rangle = a_1(0) |1 \rangle + a_2(0) |2 \rangle + a_3(0) |3 \rangle$, where $a_n(0) \in \mathbb{R}^+$ for $n \in \{1,2,3\}$, then at any time $| \psi(t) \rangle = U_r(t) | \psi(0) \rangle$, the condition $\langle n | \psi(t) \rangle \in \mathbb{R}^+$ is satisfied. Owing to their positivity, the unitary operators $U_r(t)$ correspond to rotations in a three-dimensional Euclidean space \cite{ReverseEngStirap17}. A 3D vector rotation is described by  Rodrigues' rotation formula:
\begin{equation}
\label{eq:RodriguesRotationFormula}
\mathbf{v}'= \mathbf{v} \cos \theta+ \mathbf{u} (\mathbf{u} \cdot \mathbf{v}) (1-\cos \theta) + (\mathbf{u} \times \mathbf{v}) \sin \theta \,,
\end{equation}
where $\mathbf{v}$ is the vector to be rotated, $\mathbf{v}'$ is its rotated image, $\mathbf{u}$ is the unit vector representing the axis of rotation, and $\theta$ is the angle of rotation. The corresponding relation for the quantum evolution operator is given by
\begin{equation}
\label{eq:decompositionUnophase}
U(t)=U_r(t)=\cos \theta \mathbb{I}+(1-\cos \theta ) | u \rangle \langle u| + V \sin \theta
\end{equation}
where we have introduced $|u \rangle = \sum_m u_m |m \rangle$ -  $\mathbf{u}$ is a vector with real-valued positive coefficients and of unit norm - and $V= \sum_{i,j,k} \epsilon_{ijk} u_i |k \rangle \langle j|$, with $\epsilon_{ijk}$ the Levi-Civita symbol. Any evolution operator $U_r(t)$ that leaves invariant the subset $\mathcal{S}$, that is such that $U_r(t)(\mathcal{S}) \subset \mathcal{S}$, can be expressed in this form.

Using this formalism, we investigate the saturation of the QSL. One of the conditions in Eq.~\eqref{eq:planecondition0} indicates that the three operators ${ U_0, U(t), H(t)U(t) = i\dot{U}(t) }$ must lie within a common plane. Expressing $H(t)U(t)$ as $H(t)U(t) = \lambda_1(t)\mathbb{I} + \lambda_2(t)U(t)$ with $(\lambda_1(t), \lambda_2(t)) \in \mathbb{C}^2$, we derive an alternative expression for the Hamiltonian: $H(t) =\lambda_1(t)U^\dagger(t)+ \lambda_2(t)\mathbb{I} $. Setting $a_{1,2}(t) = {\rm Re}[\lambda_{1,2}(t)] $ and $b_{1,2}(t) = {\rm Im}[\lambda_{1,2}(t)],$ the Hermiticity condition of the Hamiltonian implies that $a_1(t) (U(t)-U^{\dagger}(t))- i \: b_1(t)(U(t)+U^{\dagger}(t))= 2 i b_2(t) \mathbb{I}.$ Using Eq.~\eqref{eq:decompositionUnophase} jointly with the relation $V^\dagger = -V,$ one finds $a_1(t)  \sin \theta V - i b_1(t) \cos \theta \mathbb{I} - i b_1(t) (1- \cos \theta) | u \rangle \langle u | = i b_2(t) \mathbb{I}.$ Clearly, by linear independence of the set $\{ \mathbb{I}, | u \rangle \langle u |, V \} $, one has either $\theta(t)=0 [2 \pi]$ or $a_1(t)=b_1(t)=b_2(t)=0.$ In the first case, no motion occurs. In the second case the Hamiltonian $H(t)= a_2(t) \mathbb{I}$ merely yields a global phase factor, and the property $U_r(t)(\mathcal{S}) \subset \mathcal{S}$ further imposes $a_2(t)=0$ at all times. 

Therefore, it is impossible to saturate the QSL in a three-level system when the evolution operators are restricted to the family ${ U_r(t) }$, corresponding to unitary matrices with positive real-valued coefficients. Here, we illustrate that the bound may be unattainable. In this context, it is valuable to have a method for quantifying and optimizing the distance to speed optimality, as detailed in Sec.~\ref{SecOCT}.

\subsection{Quantum Brachistochrone for unitary evolution and the MT-QSL limit}

For state-to-state transformations, an evolution under the quantum brachistochrone does not necessarily saturate the QSL~\cite{QSLBrachistochrone23}.  
Hereafter, we investigate the relationship between time-optimal unitary evolution~\cite{Carlini07} and the framework of the operatorial quantum brachistochrone~\cite{Carlini06}.

In this framework, the Fubini-Study metric for quantum operators is utilized. It is defined as $d s_U = \langle d U, (1 - P_U) dU \rangle$, where the projection operator $P_A(X) = \langle A, X \rangle A$ is introduced for any operators $A$ and $X$. For time-independent Hamiltonians, the condition for the geodesic simplifies to:~\cite{Carlini07} 
\begin{equation}
\frac {d} {dt} \left[ (1-P_1) \left( \frac {d U} {dt} U^{\dagger} \right) \right]=0
\end{equation}
with $P_1(X)= \langle \mathbb{I}, X \rangle \mathbb{I}= \frac {1} {d} {\rm Tr}[X] \mathbb{I}.$ This condition can be interpreted as the existence of a time-independent operator $M_0$ and a time-dependent coefficient $\lambda(t)$ such that
\begin{equation}
\label{eq:brachistochronecondition}
 \dot U   = M_0 U(t) + \lambda(t) U (t).
\end{equation}
Indeed, this condition simply corresponds to a time-independent Schr\"odinger equation and is less stringent than the plane condition that arises for the MT-QSL bound, which imposes that $\dot U   = \lambda_1(t) \mathbb{I} + \lambda_2(t) U (t)$.

It is important to note that a time-optimal driving in the brachistochrone framework~\cite{Carlini07} does not necessarily saturate the  MT-QSL bound~\eqref{eq:QSL}. To show this, we discuss below a few examples of time-optimal quantum gates taken from Ref.~\cite{Carlini07}, and compare their duration with the minimal time from the MT-QSL bound~\eqref{eq:QSL}.

Consider the implementation of four-state quantum gates corresponding to
two interacting spins $1/2$, which dynamics follow the Hamiltonian $H(t)= - \sum_{j=x,y,z} J_j(t) \sigma_j^1 \sigma_j^2+ \sum_{a=1,2} B_a(t) \sigma_z^a$ with $\sigma^1_j:= \sigma_1 \otimes \mathbf{1}_2,$ $\sigma^2_j:=  \mathbf{1}_2 \otimes \sigma_j$ where $\sigma_j$ are the usual Pauli matrices for $j=x,y,z$, and the parameters $J_j(t),B_a(t)$ are free parameters determined by the brachistochrone equation. The time-optimal realization of a SWAP gate
\begin{equation}
U_{\rm SWAP} =  \left( \begin{array}{cccc} 1 & 0 & 0 & 0 \\
                                           0 & 0 & 1 & 0 \\
                                           0 & 1 & 0 & 0 \\
                                           0 & 0 & 0 & 1 \end {array}\right)
\label{eq:SWAP}                                         
\end{equation}
is reached with the time-independent Hamiltonian~\cite{Carlini07} $H_{\rm SWAP} = (\hbar \omega /\sqrt{3})[ |1\rangle\langle 1|$ + $|4\rangle\langle 4|-|2\rangle\langle 2| -|3\rangle\langle 3|+2 (| 2 \rangle \langle 3| +  | 3 \rangle \langle 2|) ]$ such that $\Delta H_{\rm SWAP}  = \hbar \omega.$ As $\langle \mathbb{I}, U_{\rm SWAP} \rangle = 1/2,$ one has $\Delta \phi=\phi(t_f)-\phi(0)= \pi/3$. The MT-QSL bound yields a time $T_{QSL}= \pi/ (3 \omega),$ which is smaller than the brachistochrone time~\cite{Carlini07} $T_{BRA}=\sqrt{3} \pi / (4 \omega) \simeq 0.433 \pi / \omega$. Similarly, when considering the ``QFT'' gate~(related to the quantum Fourier transform gate $U_{\rm QFT}$ with Hadamard gates~(see~\cite{NielsenAndChuang})),
\begin{equation}
\label{eq:QFT}
U_{\rm ``QFT''} =  \left( \begin{array}{cccc} 1 & 0 & 0 & 0 \\
                                           0 & 0 & 1 & 0 \\
                                           0 & 1 & 0 & 0 \\
                                           0 & 0 & 0 & i \end {array}\right)
\end{equation}
we obtain $|\langle \mathbb{I}, U_{\rm QFT} \rangle| = 1/\sqrt{8}$ yielding a MT-QSL bound  $T_{QSL}=\Delta \phi/ \omega \simeq 0.385 \pi/\omega$. The time corresponding to the brachistochrone~\cite{Carlini07} is slightly larger, $T_{BRA }=\sqrt{11} \pi/ (8 \omega) \simeq 0.415 \pi/\omega$. These results demonstrate that determining the path with maximum QSL efficiency is a challenging task, as the departure from the QSL typically depends on the constraints imposed on the Hamiltonian as well as on the chosen target quantum gate.

\section{Quantum optimal driving near the MT-quantum speed limit: geometric optimization with Pontryagin Maximum's principle}
\label{SecOCT}

To approach the quantum speed limit for unitary operators, we employ the Pontryagin Maximum Principle~(PMP)~\cite{PontryaginBook} using a cost functional that incorporates the conditions outlined in Eq.~\eqref{eq:twoQSLconditions}. The Pontryagin principle offers a practical optimization framework to minimize the distance of the time derivative  $\dot{U}(t)$ from the plane spanned by the initial evolution operator $U_0=\mathbb{I}$ and the current quantum evolution operator $U(t)$. We consider quantum protocols of fixed total duration $t_f$.

\subsection{Vaidman conditions and definition of the cost functionals}

To quantify the departure from the QSL, we begin with Eq.~\eqref{eq:imperfectvelocity}. We define two functions: $f_{\rm Plane}(U(t),H(t)) = ||U_{\perp \perp}(t)||^2/ \sin^2 \phi$ and $f_{\rm Phase}(U(t),H(t))= \cos^2 \beta(t),$ where $\beta(t)= {\rm Arg}[\langle U(t), \mathbb{I} \rangle \langle \mathbb{I}, U_{\perp}(t) \rangle]$. With these definitions, the instantaneous quantum speed, as given by Eq.~\eqref{eq:imperfectvelocity}, can be recast as:
\begin{equation}
\label{eq:imperfectvelocity2}
|\dot{\phi}(t)| = \frac {\Delta H} {\hbar} \sqrt{ 1 - f_{\rm Plane}(U,H)} \sqrt{ 1 - f_{\rm Phase}(U,H)}.
\end{equation}
For lighter notation, we have omitted the explicit time dependence on the right-hand side. 
These functions can be expressed as:
\begin{eqnarray}
f_{\rm Plane}(U(t),H(t)) & = &  \frac {1-|\langle \mathbb{I},U(t) \rangle|^2 -|\langle \mathbb{I},U_{\perp}(t) \rangle|^2} {1-|\langle \mathbb{I},U(t) \rangle|^2},  \label{eq:planecondition} \\
f_{\rm Phase}(U(t),H(t))  & = &  \frac {{\rm Re} \left[  \langle  U(t),  \mathbb{I} \rangle  \langle \mathbb{I}, U_{\perp}(t) \rangle  \right]^2} { |\langle  U(t),  \mathbb{I} \rangle |^2 | \langle \mathbb{I}, U_{\perp}(t) \rangle|^2}, \label{eq:phasecondition} 
\end{eqnarray}
with $U_{\perp}(t)$ defined from Eq.\eqref{eq:Uperpdefinition}. The expression holds as long as $|\langle \mathbb{I},U(t) \rangle| \neq 1$ and $ |\langle  U(t),  \mathbb{I} \rangle | | \langle \mathbb{I}, U_{\perp}(t) \rangle| \neq 0$.
The saturation of the QSL bound corresponds to $f_{\rm Plane}(U(t),H(t))=0$ and $f_{\rm Phase}(U(t),H(t))=0$ at all times.
While both conditions can be simultaneously and exactly satisfied at all times in two-dimensional quantum systems, achieving an exact solution in higher-dimensional systems depends on the available control parameters in the Hamiltonian. In such cases, a trade-off might be necessary to approximate the QSL bound by minimizing both $f_{\rm Plane}$ and $f_{\rm Phase}$.  Identifying the corresponding quantum path is the primary objective of the optimization algorithm described below.

The goal of our quantum protocols is to achieve a target evolution operator under conditions that maximize the proximity of the Quantum Speed Limit (QSL) to its theoretical bound. To this end, we employ the optimal control formalism adapted to quantum systems \cite{Ansel_2024}. It is important to note that this approach guarantees convergence to a local minimum but does not ensure that this minimum is the global one. In the following, we quantitatively assess the distance to the desired final quantum evolution operator by introducing the infidelity as terminal cost function $\Phi(U(t_f),t_f)= 1 - |\langle U_{\rm target}, U(t_f) \rangle|^2$, which decreases as the final evolution operator gets closer to the target. Finally, to incorporate the various constraints, we define the global cost functional as follows:
\begin{eqnarray}
\label{eq:costfunction}
J[U,H]  &   =  & \Phi(U(t_f),t_f) \nonumber \\
 & + & \frac {1} {t_f} \int_0^{t_f} dt' f_0(U(t'),H(t'),t')
\end{eqnarray}

The integral over the running cost $f_0(U(t),H(t),t)=p_1 f_{\rm Plane}(U(t),H(t),t)+p_2 f_{\rm Phase}(U(t),H(t),t)$ yields the QSL contribution to the cost functional $J[U,H]$. Detailed expressions for the running cost functions $f_{\rm Plane}(U(t),H(t),t)$ and $f_{\rm Phase}(U(t),H(t),t)$, in terms of the real and imaginary parts of the evolution operator and Hamiltonian, are provided in Appendix A.

The cost functional $J[U,H]$ includes two continuous components linked to the QSL conditions, namely $J_{\rm plane}[U,H]= \frac {1} {t_f} \int_0^{t_f} f_{\rm plane}(U(t),H(t)) dt$ and $J_{\rm phase}[U,H]= \frac {1} {t_f} \int_0^{t_f} f_{\rm phase}(U(t),H(t)) dt$. Their respective weights $p_1$ and $p_2$ represent a trade-off between achieving QSL saturation and obtaining a quantum operator close to the target. For $(p_1,p_2)=(0,0),$ the QSL optimization is completely deactivated, and the protocol is expected to deliver a final quantum gate with maximal target fidelity. The plane and phase running cost functions $f_{\rm Plane}(U(t),H(t))$ and $f_{\rm Phase}(U(t),H(t))$ play a symmetric role in Eq.~\eqref{eq:imperfectvelocity2}. A small mismatch in either the plane or phase condition ($f_{\rm Plane}(U,H)=\delta$ or $f_{\rm Phase}(U,H)=\delta$) results in an identical degradation of the instantaneous quantum speed. This suggests that optimal results are obtained when equal weights ($p_1=p_2$) are attributed to the two cost functionals $J_{\rm Plane}[U,H]$ and $ J_{\rm Phase}[U,H]$.

Identifying the set of quantum gates connected to the identity along a quantum path that closely follows the QSL (e.g., gates accessible via a protocol maintaining an average quantum speed above a certain percentage of the QSL) presents a significant challenge. Our optimization method offers a numerical approach to address this complex problem.

\subsection{Equations of motion in the Pontryagin formalism}

To apply the Pontryagin formalism, we introduce a real-valued matrix, $\mathbf{X}(t)$, associated with the dynamical variable—in this case, the unitary evolution matrix $U(t)$. From now on, we denote:
\begin{equation}
\mathbf{X}(t)= \left( \begin{array}{c} U_R(t) \\ U_I(t) \\ \end{array} \right)
\end{equation}
$\mathbf{X}(t)$ is a $2N \times N$ real-valued matrix and where we have decomposed the unitary evolution operator into its real and imaginary components as $U(t)=U_R(t)+ i U_I(t)$. Similarly, we write the Hamiltonian as $H(t)=H_R(t)+ i H_I(t)$.

We denote by $\mathbf{x}(t)$ the corresponding $2N^2$-dimensional column vector, obtained by arranging the matrix elements into a single column. The elements are ordered alphabetically, with the real parts listed first. For instance, in a 3-dimensional Hilbert space, $\mathbf{x}(t) = (U_{R 11}(t), U_{R 12}(t), ..., U_{R 33}(t), U_{I 11}(t), U_{I 12}(t), ..., U_{I 33}(t))^T$.

Due to the hermiticity of the Hamiltonian $H(t)$, $H_R(t)$ is defined by at most $N(N+1)/2$ independent real parameters, while $H_I(t)$ is determined by at most $N(N-1)/2$ independent parameters. The vector $\mathbf{v}(t)$ serves as a control parameter that defines $H(t)$ as a function $H(\mathbf{v}(t))$. In scenarios where all elements of the Hamiltonian can be independently controlled, $\mathbf{v}(t)$ is $N^2$-dimensional. However, in most experimental setups where only a subset of Hamiltonian couplings can be independently tuned, the dimension of $\mathbf{v}(t)$ is smaller.

The Schrödinger equation, $i \hbar \dot{U}(t) = H(\mathbf{v}(t)) U(t)$, can be reformulated in terms of real-valued matrices as:
\begin{equation}
\dot{\mathbf{X}}(t)= \left( \begin{array}{c} \dot{U}_R  \\  \dot{U}_I \end{array} \right) = \frac {1} {\hbar} \left( \begin{array}{cc} H_I(\mathbf{v}(t)) & H_R(\mathbf{v}(t)) \\
- H_R(\mathbf{v}(t)) &  H_I(\mathbf{v}(t)) \end{array} \right) \mathbf{X}(t)
\end{equation}
or, more compactly, as a single real-valued vectorial differential equation
\begin{equation}
\dot{\mathbf{x}}(t)= f(\mathbf{x}(t),\mathbf{v}(t),t),
\end{equation}
governed by a vectorial control $\mathbf{v}(t)$. This provides a suitable framework for applying Pontryagin's Maximum Principle. Note that in this context, $f(\mathbf{x}(t), \mathbf{v}(t), t)$ is a linear function of $\mathbf{x}(t)$, which can be expressed as $f(\mathbf{x}(t), \mathbf{v}(t), t) = M[\mathbf{v}(t)] \mathbf{x}(t)$, where $M[\mathbf{v}(t)]$ is a matrix determined by the Hamiltonian $H(t)$.
 
The Pontryagin Hamiltonian is given by
\begin{eqnarray}
\label{eq:Pontryagin}
H_P(\mathbf{x}(t),\mathbf{p}(t),\mathbf{v}(t),t) & = & - \frac {1} {t_f} f_0(\mathbf{x}(t),\mathbf{v}(t),t) \nonumber \\
& + & \mathbf{p}(t) \cdot  f(\mathbf{x}(t),\mathbf{v}(t),t),
\end{eqnarray}
where we have introduced the adjoint vector $\mathbf{p}(t)$. The vector $\mathbf{x}(t)$ and its adjoint $\mathbf{p}(t)$ fulfill by construction the Hamilton equations:
\begin{equation}
    \label{eq:Hamiltoneqs}
\dot{\mathbf{p}}= - \frac {\partial H_P } {\partial \mathbf{x}},   \qquad \qquad \dot{\mathbf{x}}= \frac {\partial H_P } {\partial \mathbf{p}} .
\end{equation}
and the boundary condition (for fixed final time $t_f$)
\begin{equation}
\label{eq:boundarycondition}
\mathbf{p}(t_f) = \left. - \frac {\partial \Phi(\mathbf{x}(t),t)} {\partial \mathbf{x}} \right|_{t=t_f}.
\end{equation}
The PMP states that the optimal control parameter $\mathbf{v}^*(t)$ is obtained by maximizing the Pontryagin Hamiltonian at any time $t$ along the trajectory $(\mathbf{x}(t),\mathbf{p}(t))$:
\begin{equation}
\left. \frac {\partial H_P (\mathbf{x}(t),\mathbf{p}(t),\mathbf{v}(t),t)} {\partial \mathbf{v}} \right|_{\mathbf{v}(t)=\mathbf{v}^*(t)} =0 .
\end{equation}

Let us now express the dynamical equation for the adjoint $\mathbf{p}(t)$ in a more explicit form. Following the same correspondence between the evolution operator $U(t)$ and its associated column vector $\mathbf{x}(t)$, the column vector $\mathbf{p}(t)$ can be associated with a complex-valued adjoint operator $U^P(t)$. It is important to emphasize that this adjoint operator has no physical significance and is not necessarily unitary—it is merely a mathematical object arising from the optimization method. From Eqs.~\eqref{eq:Hamiltoneqs}, this operator satisfies the dynamical equation~(see Appendix B):
\begin{eqnarray}
\label{eq:adjointpropagationequation}
\dot{U}^P(t) & = &  \frac {1} {t_f} \bigg( \frac {\partial f_{0}(U(t),H(\mathbf{v}(t)))} {\partial U_R}+ i \: \frac {\partial f_{0}(U(t),H(\mathbf{v}(t)))} {\partial U_I} \bigg) \nonumber \\ & -&   
   \frac {i} {\hbar} H(\mathbf{v}(t)) U^P(t)  .
\end{eqnarray}
When compared to the dynamics of the unitary operator $U(t)$, the adjoint operator includes additional driving terms associated with the derivatives of the cost functionals. Finally, we can express the Pontryagin Hamiltonian and the boundary conditions for the adjoint as:
\begin{eqnarray}
\label{eq:Pontryagin2}
H_P(\mathbf{x}(t),\mathbf{p}(t),t) & = & - \frac  {1} {t_f} f_0(U(t),H(\mathbf{v}(t)),t)     \\
& \: &  + \frac {d} {\hbar}  {\rm Re}[ \langle U^P(t), H(\mathbf{v}(t)) U(t) \rangle ]  \nonumber
\end{eqnarray}
with 
\begin{equation}
U^P(t_f)  =   \frac {2} {d} \langle U_{\rm target}, U(t_f) \rangle U_{\rm target}.
\label{eq:boundaryconditionadjointoperator}    
\end{equation}
With the dynamics of the quantum operator and its adjoint at hand, one can apply a quantum optimal control procedure similar to that described in Refs.~\cite{Sugny21,DGOPRX21}. The time evolution over the interval $[0, t_f]$ is discretized into $N_t$ time bins, each centered at $t_m$ for $1 \leq m \leq N_t$ (in our numerical examples, $N_t = 200$). Assuming $M$ independent driving parameters in the Hamiltonian, the time-dependent control parameter $[\mathbf{v}(t)]$ is represented as a $N_t \times M$-dimensional vector, $[\mathbf{v}(t)] = [v_1(t), ..., v_M(t)]$.

Starting with an initial guess for the control parameter $[\mathbf{v}^{(0)}(t)]$—typically chosen as constant—we proceed iteratively as follows. For a given (generally time-dependent) control parameter $[\mathbf{v}^{(n)}(t)]$, we numerically propagate the evolution operator $U^{(n)}(t)$ over the interval $[t, t_f]$. The boundary condition~\eqref{eq:boundaryconditionadjointoperator} provides the final value of the adjoint operator $U^{P (n)}(t_f)$.

Next, using Eq.~\eqref{eq:adjointpropagationequation}, we perform a numerical back-propagation to determine the adjoint operator on the interval $[0, t_f]$. At each discrete time $t_m$, the Pontryagin Hamiltonian~\eqref{eq:Pontryagin2} and its gradient with respect to the control parameter are computed. Finally, we update the control parameter as:
\begin{eqnarray}
&&\mathbf{v}^{(n+1)}(t_m)  =  \mathbf{v}^{(n)}(t_m)  \\
&& +  \epsilon \:  \nabla_{\mathbf{v}} H_P(U^{(n)}(t_m),U^{P (n)}(t_m),\mathbf{v},t_m))|_{\mathbf{v}=\mathbf{v}^{(n)}(t_m)} \nonumber
\end{eqnarray}
where the parameter $\epsilon > 0$ is chosen to be sufficiently small. The procedure is then iterated until a satisfactory solution is reached. With an appropriate choice of $\epsilon$, the cost functional should decrease over successive iterations. Alternatively, one can use adaptive values $\epsilon \equiv \epsilon^{(n)}$, which are reduced whenever the cost functional~\eqref{eq:costfunction} fails to decrease during a given iteration.

\section{Quantum Speed Limit Optimization in a 4-dimensional space}
\label{Secexamples}

In this section, we apply our optimization method in a four-dimensional space. Using the cost functionals~(\ref{eq:costfunction}), we compare two optimization methods: the PMP optimization described in the previous section, and the Chopped Random Basis quantum optimization~(CRAB)~\cite{CRAB11}.

 We investigate the implementation of the following set of four 2-qubit target gates: the quantum Fourier transform (QFT) gate ($U_{\rm QFT}$), the gate $U_{\rm ``QFT''}$ (defined in Eq.~\eqref{eq:QFT}), the Hadamard gate ($U_{\rm Hadamard}$), and the CNOT gate ($U_{\rm CNOT}$), which is equivalent to the SWAP gate (Eq.~\eqref{eq:SWAP}) up to a label permutation. We recall the expressions for the QFT and Hadamard gates:   
\begin{eqnarray}
U_{\rm QFT} =  \frac {1} {2} & &  \left( \begin{array}{cccc} 1 & 1 & 1 & 1 \\
                                           1 & i & -1 & -i \\
                                           1 & -1 & 1 & -1 \\
                                           1 & -i & -1 & i \end {array}\right), \nonumber \nonumber \\
                                           U_{\rm Hadamard}  = \frac {1} {2} & &  \left( \begin{array}{cccc} 1 & 1 & 1 & 1 \\
                                           1 & -1 & 1 & -1 \\
                                           1 & 1 & -1 & -1 \\
                                           1 & -1 & -1 & 1 \end {array}\right).
\end{eqnarray}
We target a minimum fidelity of $99 \%$ using our metric. In practice, however, our algorithm yields an even more accurate implementation of these gates (up to a global constant phase).

\subsection{Optimization in a 4-dimensional space with the Pontryagin Maximum Principle}

We consider an evolution governed by the Hamiltonian $H(t) =  \sum_{j=1}^4 \sum_{i<j} \left[ \hbar \left( \Omega^R_{ij}(t)+i \: \Omega^I_{ij}(t) \right) | j \rangle \langle i | + h.c. \right]+ \sum_{i=1}^4  \hbar \Delta_{ii}(t) | i \rangle \langle i|$ over a total time $t_f$. We initiate the optimization routine with a constant guess Hamiltonian, setting all available couplings to a small constant value $\Omega^{R,I}_{ij}(t)=\Delta_{ii}(t)\equiv 0.01/t_f$ for $0 \leq t \leq t_f$. The optimization evolves over typically 200 iterations of our routine. The protocol's performance is evaluated at each iteration by numerically solving the time-dependent Schrödinger equation. Special care must be taken since our optimal control protocol simultaneously affects the instantaneous quantum speed, $\dot{\phi}(t)$, and its upper bound, $\Delta H(t)$. The latter varies over time for the considered time-dependent Hamiltonians. This is why we use the ratio of these quantities, namely the QSL efficiency ($\eta(t)= \hbar \dot{\phi} / \Delta H$), as our measure of the instantaneous quantum speed's optimality. In the examples below, we use the average QSL efficiency as our figure of merit $\overline{\eta}= \frac {1} {t_f} \int_0^{t_f} \eta(t) dt$. As discussed hereafter, the average quantum speed depends on the chosen cost functional and on the available control parameters. 

 Here, we assume full control over all Hamiltonian parameters, i.e. 16-dimensional control vector $\mathbf{v}(t)=(\Delta_{ii}(t),\Omega_{ij}^R(t),\Omega_{ij}^I(t))$, ordered alphabetically  for $(i,j) \in \{1,2,3,4\}^2$ and $i<j$.  As a benchmark for our method, we first illustrate an example of PMP optimization focusing solely on the fidelity to the target gate, that is with weights set as $(p_1 = p_2 = 0)$. In this case, the routine achieves perfect fidelity ($1-\mathcal{F} \leq  10^{-9}$). In Table~\ref{tab:OptFidelityOnly}, we present the plane and phase cost functionals $J_{\rm plane}[U,H]$  and $J_{\rm phase}[U,H]$~(defined previously as the time-average of $f_{\rm plane}(U(t),H(t))$ and $f_{\rm phase}(U(t),H(t))$ along the quantum trajectory), along with the average quantum speed efficiencies obtained for the set of quantum gates considered. The optimization routine outputs the time-dependent couplings as a vector discretized over a time grid of $N_t=10^3$ time bins, which is sufficient for effective optimization. Immediately after the optimization, we add a stage to smooth the resulting pulses by averaging over blocks of $N_{\rm av}=20$ time bins. Finally, to obtain more accurate QSL and quantum fidelity estimates, we perform a linear interpolation of the optimized pulses, followed by a numerical resolution of the Schrödinger equation with a smaller time step $dt=10^{-4} \: t_f$. Similarly to quantum states, a perfect fidelity implies that the target gate is achieved up to a global phase factor $\varphi={\rm Arg}[\langle U_{\rm target}, U(t_f) \rangle]$. This extra phase can be removed by adding a final stage where a global energy shift $H=\hbar \Delta \omega \mathbb{I}$ (associated to a null QSL as $\Delta H=0$) is applied during $\tau=| \varphi/ \Delta \omega|$~\cite{FootnoteQSL1}.

 \begin{table}[htbp]
    \centering
    \begin{tabular}{|l|c|c|c|c|c|}
        \toprule
        \textbf{Gate} & \textbf{Fidelity} &   $J_{\rm Plane}[U,H]$ & $J_{\rm Phase}[U,H]$ & \:  $\overline{\eta}$ \:    \\
       \hline
         $U_{\rm QFT}$     & 1.000    & 0.076 & 0.080             &  0.922                  \\
        $U_{\rm ``QFT''}$  & 1.000 & 0.063 & 0.079               & 0.928                   \\
           $U_{\rm Hadamard}$   &  1.000 &  0.035 &      0.000           &         0.982              \\ 
           $U_{\rm CNOT}$   &  1.000       & $10^{-4}$ & 0.334           & 0.769                     \\
        \hline
    \end{tabular}
\caption{Implementation of quantum gates by fidelity optimization only ($p_1=p_2=0$). Other numerical parameters: $200$ iterations, $\epsilon=0.25/t_f$, discretization in $N_t=1000$ time bins.}
\label{tab:OptFidelityOnly}
\end{table}
Table~\ref{tab:OptFidelityOnly} confirms that higher values for either cost functional ($J_{\rm Plane}[U,H]$ or $J_{\rm Phase}[U,H]$) lead to lower quantum speed efficiencies.
Depending on the target gate, optimization focused solely on fidelity may result in a trajectory where one of the functionals is negligible. For example, with the Hadamard gate ($U_{\rm Hadamard}$), the phase functional effectively cancels out (up to numerical noise). Conversely, for the CNOT gate ($U_{\rm CNOT}$), the $J_{\rm Plane}[U,H]$ functional is negligible, although $J_{\rm Phase}[U,H]$ remains high, significantly degrading the quantum speed efficiency. 

These results suggest that the ideal weights $(p_1,p_2)$ for achieving the best trade-offs between quantum speed and quantum fidelity depend on the specific target gate. For example, an optimization by minimizing only the phase cost functional ($p_1=0, p_2=0.5$) is expected to be ineffective for the Hadamard gate. This is because its ``natural" path (i.e., the path obtained without influence from the QSL functionals) already yields a negligible phase cost, $J_{\rm Phase}[U,H]$. Conversely, such an optimization is expected to significantly improve the implementation of the CNOT gate, which otherwise suffers from a high phase cost in the absence of QSL functionals.

First, we optimized by minimizing only one cost functional at a time, while maintaining the cost associated with infidelity. Specifically, we set $p_1=0.5$ and $p_2=0$, which directed the optimization routine to minimize the plane cost functional $J_{\rm Plane}[U,H]$, while disregarding the phase cost functional $J_{\rm Phase}[U,H]$. 

The results, presented in Table~\ref{tab:OptFidelityPlane}, show that the gate fidelities remain higher than $99.8\%$. Furthermore, the procedure considerably reduced the quantum speed inefficiency ($\delta \overline{\eta}=1-\overline{\eta}$) when compared to the results in Table~\ref{tab:OptFidelityOnly}. The only exception is the CNOT gate, for which only a marginal improvement was obtained. For this gate, the fidelity optimization in Table~\ref{tab:OptFidelityOnly} already yielded a path with a very small plane cost. 

Notably, the Hadamard gate, which was previously implemented with $\overline{\eta}=98.2\%$ of the QSL, is now implemented with no loss in quantum fidelity at $\overline{\eta}=99.99\%$ of the QSL, thus extremely close to the saturation limit. As expected, the plane cost functional $J_{\rm Plane}[U,H]$ is reduced compared to its values in Table~\ref{tab:OptFidelityOnly}. Interestingly, the phase cost functional $J_{\rm Phase}[U,H]$ is also lower than in the previous case, even though we did not explicitly minimize its corresponding functional here. We attribute this side benefit to the fact that minimizing $J_{\rm Plane}[U,H]$ already drives the system closer to the optimal QSL path, where both functionals are minimal.

\begin{table}[htbp]
    \centering
    \begin{tabular}{|l|c|c|c|c|}
        \toprule
        \textbf{Gate} & \textbf{Fidelity} & $J_{\rm Plane}[U,H]$ & $J_{\rm Phase}[U,H]$ &  \:  $\overline{\eta}$ \:   \\
       \hline
         $U_{\rm QFT}$     & 0.9996             &  0.041    &  0.045 &  0.958            \\ 
        $U_{\rm ``QFT''}$  & 0.9998            &   0.043   &  0.048  & 0.954                \\ 
           $U_{\rm Hadamard}$   &  1.000            & $1.1 \times 10^{-4}$   & 0   &  0.9999                \\ 
           $U_{\rm CNOT}$   & 1.000          &  $1.4 \times 10^{-4}$   & 0.327  &  0.775                      \\ 
        \hline
    \end{tabular}
\caption{Fidelity optimization while minimizing only the plane cost functional ($p_1=0.5, p_2=0$) with $200$ iterations. $100$ iterations, variable step (initially $\epsilon=0.25/t_f$, increased by $50\%$ every 15 iterations, decreased by $10\%$ in case of cost increase) Other numerical parameters are identical to Table~\ref{tab:OptFidelityOnly}.} 
\label{tab:OptFidelityPlane}
\end{table}

Conversely, Table \ref{tab:OptFidelityPhase} presents the results of a fidelity optimization while minimizing only the phase cost functional $J_{\rm Phase}[U,H]$, using therefore weights $(p_1=0,p_2=0.5)$.

When compared to the results of the fidelity optimization  while minimizing only the plane cost functional shown in Table \ref{tab:OptFidelityPlane}, we observe that the QFT ($U_{\rm QFT}$), the  QFT-analog~($U_{\rm ``QFT''}$) and the Hadamard gates are implemented with slightly lower average QSL efficiencies. 

Nevertheless, the CNOT gate now exhibits an average QSL efficiency above 90\%, a significant improvement from the 77\% achieved with previous optimizations. This gain occurs because the ``natural" path for the CNOT gate (obtained through pure fidelity optimization) suffers from a high cost associated with the $J_{\rm Phase}[U,H]$ functional, which is substantially reduced here. This example clearly illustrates that the appropriate weights ($p_1,p_2$) for QSL optimization are gate-dependent.

\begin{table}[htbp]
     \centering
    \begin{tabular}{|l|c|c|c|c|}
        \toprule
        \textbf{Gate} & \textbf{Fidelity} & $J_{\rm Plane}[U,H]$ & $J_{\rm Phase}[U,H]$ &  \:  $\overline{\eta}$ \:  \\
       \hline
         $U_{\rm QFT}$     &  0.994           &  0.073  & 0.050   &  0.936                 \\
        $U_{\rm ``QFT''}$  &   0.994       &   0.059    &  0.048 &  0.947                  \\
           $U_{\rm Hadamard}$   &  0.996        &  0.028  & 0.000  &  0.986                \\
           $U_{\rm CNOT}$   &  0.9987          &  0.108  &  0.051  &  0.908                       \\
        \hline
    \end{tabular}
\caption{Fidelity optimization while minimizing only the phase cost functional ($p_1=0, p_2=0.5$). Variable step (initially $\epsilon=0.25/t_f$, increased by $50\%$ every 15 iterations, decreased by $10\%$ in case of cost increase). Up to $200$ iterations. Other numerical parameters are identical to Table~\ref{tab:OptFidelityOnly}.}
\label{tab:OptFidelityPhase}
\end{table}

Next, we performed the PMP fidelity optimization routine while minimizing both plane and phase cost functionals $J_{\rm plane}(U,H)$ and $J_{\rm phase}(U,H)$ by setting $p_1=p_2=0.5$. The results are summarized in Table \ref{tab:OptPlanePhase}.

This simultaneous optimization of both functionals yields the best performance in terms of QSL efficiency, outperforming the results in both Table \ref{tab:OptFidelityPlane} and Table \ref{tab:OptFidelityPhase} for nearly all gates. Indeed, at the cost of a minor degradation in quantum fidelity ($\delta \mathcal{F} \sim 0.1-0.2\%$), we achieve a gain of at least a few percent in the average QSL efficiency ($\overline{\eta}$).

For the QFT quantum gates ($U_{\rm QFT}$ and $U_{\rm ``QFT''}$), the departure from quantum speed optimality ($\delta{\overline{\eta}}=1-\overline{\eta}$) observed in Table \ref{tab:OptFidelityOnly} is reduced by approximately 50\%. This deviation is reduced by nearly two-thirds for the CNOT gate. For the Hadamard gate, this optimization performs as well as when considering the fidelity optimization involving only the plane cost functional~(Table~\ref{tab:OptFidelityPlane}). This was expected as the ``natural path'' for this gate carries no phase cost.

\begin{table}[htbp]
    \centering
    \label{tab:my_results}
    \begin{tabular}{|l|c|c|c|c|}
        \toprule
        \textbf{Gate} & \textbf{Fidelity} & $J_{\rm Plane}[U,H]$ & $J_{\rm Phase}[U,H]$ &  \: $\overline{\eta}$ \:   \\
       \hline
         $U_{\rm QFT}$     &   0.9987         & 0.034 &   0.032       &  0.967        \\
        $U_{\rm ``QFT''}$  &   0.9979          &   0.034 &   0.032             &    0.967         \\
           $U_{\rm Hadamard}$  &  1.000          &  $1.3 \times 10^{-4}$ &   0       &  0.9999             \\
           $U_{\rm CNOT}$   &   0.9989         & 0.0001 &   0.12       &  0.921                      \\
        \hline
    \end{tabular}
\caption{Fidelity optimization while minimizing both the plane and phase cost functionals with the same weight ($p_1=p_2=0.5$).  Fixed small parameter: $\epsilon=0.25/t_f$ and $200$ iterations. Other numerical parameters identical to Table~\ref{tab:OptFidelityOnly}.}
\label{tab:OptPlanePhase}
\end{table}

\begin{widetext}

\begin{figure}[t!]
\includegraphics[width=15cm]{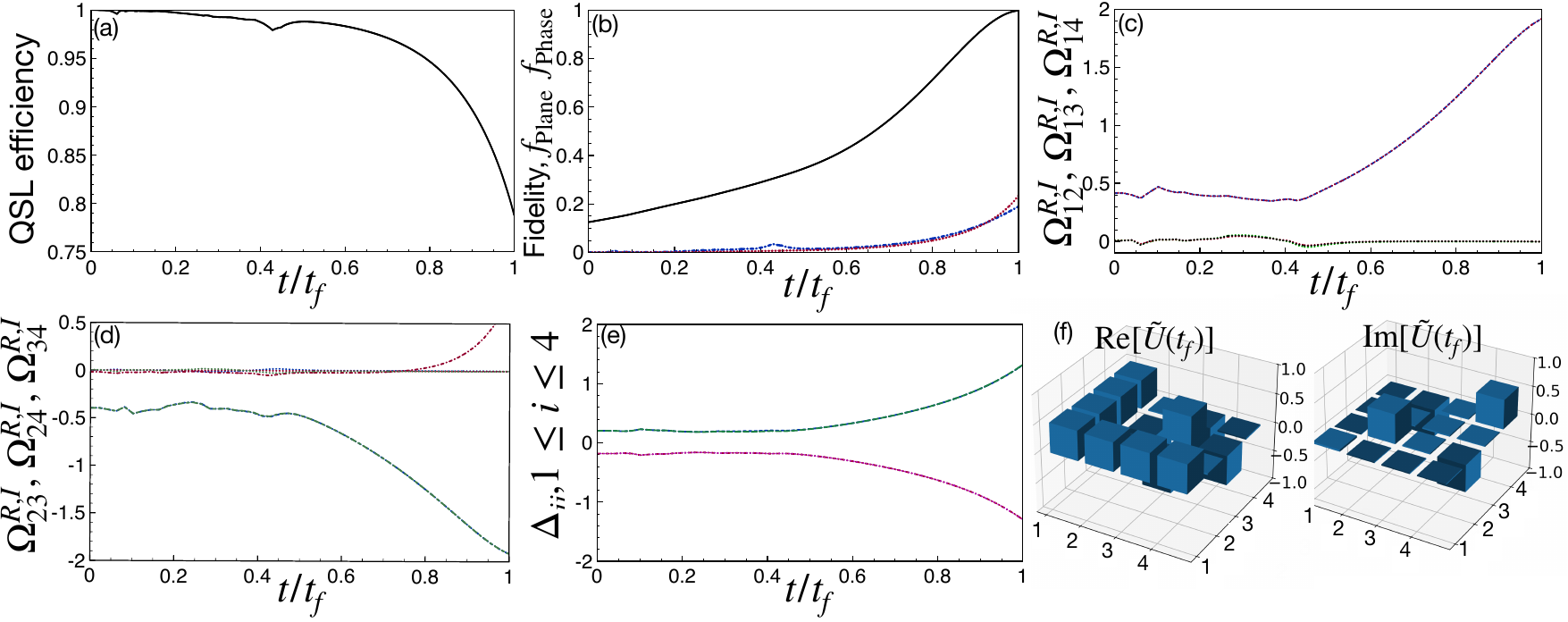}
\caption{Implementation of the gate $U_{\rm QFT}$ after geometric PMP optimization with a full Hamiltonian control [$(p_1=p_2=0.5)$, 1st line of Table~\ref{tab:OptPlanePhase}]. (a):Instantaneous QSL efficiency $\eta(t)$ as a function of the rescaled time $t/t_f$. (b):Fidelity $\mathcal{F}=|\langle U_{\rm target},U(t) \rangle|^2,$ cost functionals $f_{\rm Plane}(U(t),H(t))$~(dash-dotted blue line) and $f_{\rm Phase}(U(t),H(t))$~(dotted red line) as a function of the  rescaled time $t/t_f$. (c):Dynamical couplings $\Omega_{12}^R(t)$ (dash-dotted blue line), $\Omega_{12}^I(t)$ (dotted blue line), $\Omega_{13}^R(t)$ (dash-dotted red line), $\Omega_{13}^I(t)$ (dotted red line), $\Omega_{23}^R(t)$ (dash-dotted green line), $\Omega_{23}^I(t)$ (dotted green line). Note that the lines of $\Omega_{12}^R(t)$, $\Omega_{13}^R(t)$ and $\Omega_{14}^R(t)$ are superimposed, as well as the lines of $\Omega_{12}^I(t)$, $\Omega_{13}^I(t)$ and $\Omega_{14}^I(t)$. (d):Dynamical couplings $\Omega_{23}^R(t)$ (dash-dotted blue line), $\Omega_{23}^I(t)$ (dotted blue line), $\Omega_{24}^R(t)$ (dash-dotted red line), $\Omega_{24}^I(t)$ (dotted red line), $\Omega_{34}^R(t)$ (dash-dotted green line), $\Omega_{34}^I(t)$ (dotted green line). The lines of $\Omega_{23}^R(t)$ and $\Omega_{34}^R(t)$ are superimposed. (e):Dynamical couplings $\Delta_{11}(t)$ (dash-dotted blue line), $\Delta_{22}(t)$ (dash-dotted red line), $\Delta_{33}(t)$ (dotted green line), $\Delta_{44}(t)$ (dotted purple line). The lines of $\Delta_{11}(t)$ and $\Delta_{33}(t)$ are superimposed, as well as $\Delta_{22}(t)$ and $\Delta_{44}(t)$. All couplings in (c,d,e) are given in units of $1/t_f$. (f):Matrix elements of the operator $\tilde{U}(t_f)=U(t_f) e^{- i \varphi}$~(real and imaginary parts), with $\varphi= {\rm Arg}[\langle U_{\rm target}, U(t_f) \rangle]$.}
 \label{fig:FullQSLPicture}
\end{figure}
\end{widetext}

 For illustration, Figure \ref{fig:FullQSLPicture} provides additional details on the time-dependent evolution of the system during the optimal implementation of the $U_{\rm QFT}$ gate. This figure sketches the time-dependent profiles of the instantaneous quantum speed efficiency $\eta(t)$, the running plane and phase cost functions $f_{\rm Plane}(U(t),H(t))$ and $f_{\rm Phase}(U(t),H(t))$ and the time-dependent couplings $\{ \Omega_{ij}^{R,I}(t),\Delta_{ii}(t) \}$ involved in the gate implementation.

\begin{widetext}

\begin{figure}[t!]
\includegraphics[width=15cm]{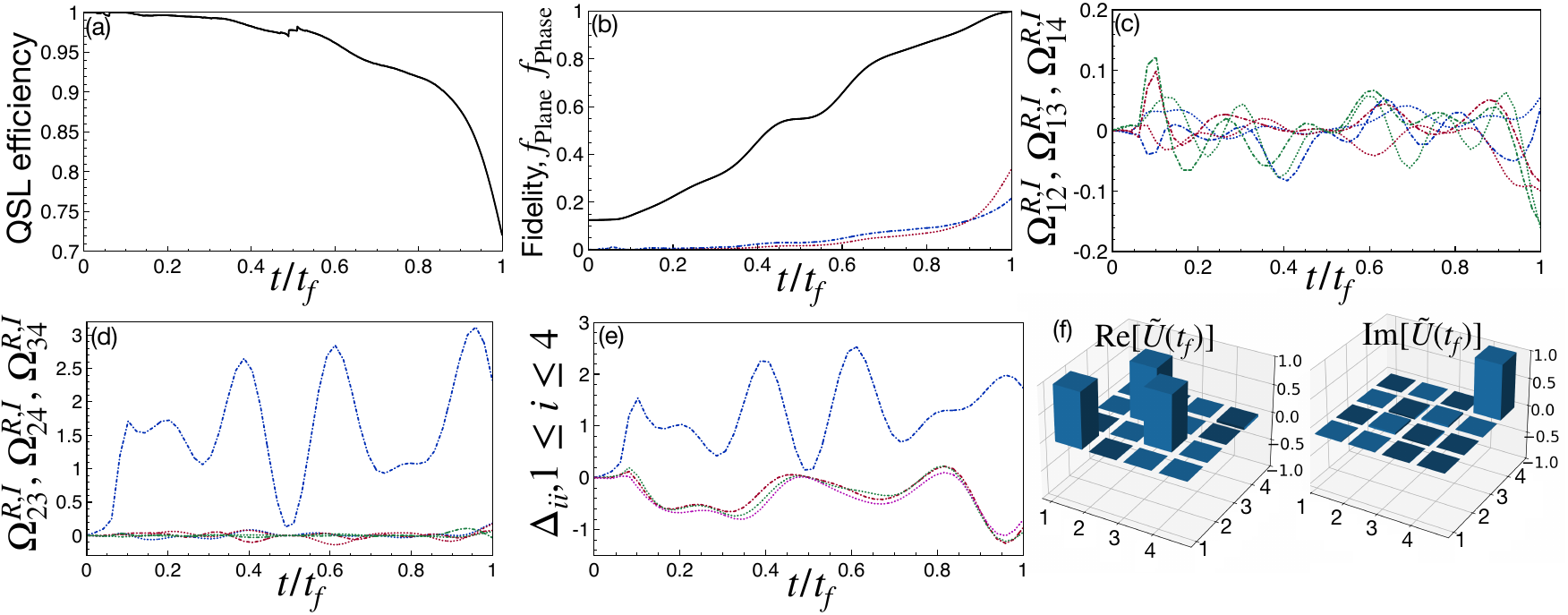}
\caption{Implementation of the gate $U_{\rm ``QFT''}$ after geometric CRAB optimization with a full Hamiltonian control~[$(p_1=0.5,p_2=0.1)$, 2nd line of Table~\ref{tab:CRABQSLOptimization}]. (a):Instantaneous QSL efficiency $\eta(t)$ as a function of the rescaled time $t/t_f$. (b):Fidelity $\mathcal{F}=|\langle U_{\rm target},U(t) \rangle|^2,$ running cost functions $f_{\rm Plane}(U(t),H(t))$~(dash-dotted blue line) and $f_{\rm Phase}(U(t),H(t))$~(dotted red line) as a function of the rescaled time $t/t_f$. (c):Dynamical couplings $\Omega_{12}^R(t)$ (dash-dotted blue line), $\Omega_{12}^I(t)$ (dotted blue line), $\Omega_{13}^R(t)$ (dash-dotted red line), $\Omega_{13}^I(t)$ (dotted red line), $\Omega_{23}^R(t)$ (dash-dotted green line), $\Omega_{23}^I(t)$ (dotted green line). 
(d):Dynamical couplings $\Omega_{23}^R(t)$ (dash-dotted blue line), $\Omega_{23}^I(t)$ (dotted blue line), $\Omega_{24}^R(t)$ (dash-dotted red line), $\Omega_{24}^I(t)$ (dotted red line), $\Omega_{34}^R(t)$ (dash-dotted green line), $\Omega_{34}^I(t)$ (dotted green line). (e): Dynamical couplings $\Delta_{11}(t)$ (dash-dotted blue line), $\Delta_{22}(t)$ (dash-dotted red line), $\Delta_{33}(t)$ (dotted green line), $\Delta_{44}(t)$ (dotted purple line). All couplings in (c,d,e) are given in units of $1/t_f$.(f):Matrix elements of the operator $\tilde{U}(t_f)=U(t_f) e^{- i \varphi}$~(Real and Imaginary parts), with $\varphi= {\rm Arg}[\langle U_{\rm target}, U(t_f) \rangle]$.}
 \label{fig:FullQSLoptimisation}
\end{figure}

\end{widetext}

As confirmed by the histogram plot in Figure \ref{fig:FullQSLoptimisation}f, the matrix elements of the final evolution operator - multiplied by a global phase factor $e^{-i \varphi}$ with $\varphi= {\rm Arg}[\langle U_{\rm target}, U(t_f) \rangle]$ - are visually identical to those of the target gate $U_{\rm QFT}$. This serves as visual confirmation that our metric ensures convergence towards the target gate.

Finally, we consider a setup with more limited access to the Hamiltonian couplings, for example, assuming that the imaginary couplings ${ \Omega_{ij}^I(t) }$ are unavailable. This corresponds to a scenario where Rabi pulses $\Omega_{ij}(t)$ can be applied between all sub-levels, but with the phase fixed to $\phi_{ij}(t) = 0$.  Besides, we also assume that only one detuning ${ \Delta_{11}(t) }$ is accessible. The 7-dimensional control vector, noted $\mathbf{v}(t)=(\Delta_{11}(t),\Omega_{ij}^R(t))$ with an alphabetic ordering for $(i,j) \in \{1,2,3,4\}^2$ and $i<j$ is initialized using  $\mathbf{v}^{(0)}(t)=v_0 (1,1,1,1,1,1,1)^T$ with $v_0= 0.01 /t_f$ for $0 \leq t \leq t_f$. Remaining control parameters are such that $\Delta_{ii}(t)=0$ for $i \geq 2$ and $\Omega_{ij}^I(t)=0$ at all times.

In this context, the optimization's performance strongly depends on the target quantum gate. We first investigated the quantum fidelity and QSL cost functionals obtained through an optimization with respect to fidelity only (Table \ref{tab:FidOnlyLimited}). Despite the limited available couplings, the optimization routine delivered very high fidelity for three of the target gates. The only exception was the CNOT gate, for which we obtained a poor fidelity of only 65\%.

The values of the cost functionals and efficiencies can differ significantly from those obtained in a full-control scenario. Both the QFT and Hadamard gates were obtained with a significantly lower QSL efficiency, following a path that yielded higher QSL cost functionals $J_{\rm Plane}[U,H]$ and $J_{\rm Phase}[U,H]$. However, limiting to seven couplings did not affect the $U_{\rm ``QFT''}$ gate, for which the optimization yielded similar quantum speed and fidelity.

Next, we proceeded to the fidelity optimization while minimizing both plane and phase cost functionals $J_{\rm Plane}[U,H]$ and $J_{\rm Phase}[U,H]$ with $p_1=p_2=0.5$. The results are summarized in Table~\ref{tab:OptPlanePhaseLimited}.

Surprisingly, our routine delivered a CNOT gate with high fidelity. We attribute this to the steepest-gradient descent exploring a different path, thus avoiding a local cost minimum previously associated with poor fidelity. The QFT, QFT analog and Hadamard gates were realized at a higher quantum speed efficiency than in the pure-fidelity optimization of Table \ref{tab:FidOnlyLimited}.

Nevertheless, the optimization yielded degraded QSL efficiencies when compared to the full-control scenario presented in Table~\ref{tab:OptPlanePhase}. The difference depends sharply on the target gate: the QFT analog gate ($U_{\rm ``QFT''}$) was implemented with similar QSL and fidelity despite the limited couplings. This contrasts sharply with the Hadamard gate, which was achieved nearly at the QSL saturation limit with complete control but is now implemented at a low QSL efficiency ($\overline{\eta} \simeq 0.74\%$). A similar result holds for the CNOT gate, implemented with a QSL efficiency lower than in the full-control optimization. Finally, the QFT gate ($U_{\rm QFT}$) is now obtained with a similar QSL efficiency but with a slightly lower fidelity than the one reported in Table~\ref{tab:OptPlanePhase}.

These results demonstrate that achieving — or even approaching — the quantum speed limit critically depends on both the target gates and the available dynamical couplings. The procedure's performance, specifically the trade-off between quantum speed and quantum fidelity, improves with the number of available control parameters, as the geometric optimization routine can then explore a broader range of quantum paths.

\begin{table}[htbp]
    \centering
    \label{tab:my_results}
    \begin{tabular}{|l|c|c|c|c|}
        \toprule
        \textbf{Gate} & \textbf{Fidelity} & $J_{\rm Plane}[U,H]$ & $J_{\rm Phase}[U,H]$ &  \: $\overline{\eta}$ \:   \\
       \hline
         $U_{\rm QFT}$     &   0.9995          &  0.426 &  0.104      &  0.700           \\
        $U_{\rm ``QFT''}$  & 1.000   &   0.063 &   0.079             &     0.928             \\
           $U_{\rm Hadamard}$   & 0.9998     & 0.507 &  0.0001   &  0.672             \\
           $U_{\rm CNOT}$   &   0.654    & 0.066 & 0.063 &  0.935                      \\
        \hline
    \end{tabular}
\caption{Implementation of quantum gates using fidelity as the sole optimization criterion ($p_1=p_2=0$) with a control limited to 7 couplings $\mathbf{v}(t)=(\Delta_{11}(t),\Omega_{ij}^R(t))$.  Other numerical parameters: $100$ iterations, variable step (initially $\epsilon=0.25/t_f$, increased by $50\%$ every 15 iterations, decreased by $10\%$ in case of cost increase), with $N_t=1000$ time bins.}
\label{tab:FidOnlyLimited}
\end{table}

\begin{table}[htbp]
    \centering
    \label{tab:my_results}
    \begin{tabular}{|l|c|c|c|c|}
        \toprule
        \textbf{Gate} & \textbf{Fidelity} & $J_{\rm Plane}[U,H]$ & $J_{\rm Phase}[U,H]$ &  \: $\overline{\eta}$ \:    \\
       \hline
         $U_{\rm QFT}$     &   0.953          &  0.0046 &   0.0035      & 0.960          \\
        $U_{\rm ``QFT''}$  & 0.9993   &   0.034 &   0.035             &    0.965           \\
           $U_{\rm Hadamard}$   & 0.9762     & 0.349 &  0.041   &  0.744           \\
           $U_{\rm CNOT}$   &   0.9974    &   0.281 & 0.011 &  0.806                      \\
        \hline
    \end{tabular}
\caption{Fidelity optimization while minimizing both the plane and phase cost functionals with the same weight ($p_1=p_2=0.5$) with a control limited to 7 couplings $\mathbf{v}(t)=(\Delta_{11}(t),\Omega_{ij}^R(t))$. Up to $200$ iterations with a variable step as in Table~\ref{tab:OptFidelityPlane} and $N_t=1000$ time bins.}
\label{tab:OptPlanePhaseLimited}
\end{table}


\subsection{Geometric Optimization with the Chopped Random Basis quantum optimization}

We now compare our geometric approach with CRAB (Chopped Randomized Basis) optimization. Differences in results are expected, as various optimization procedures can converge to different local minima.
We again consider a time-dependent, $N_C$-dimensional control parameter $\mathbf{v}(t)$ representing the Hamiltonian couplings. Here, $N_C=N^2$ in an $N$-dimensional Hilbert space when all couplings can be independently controlled.

The core idea of the CRAB method is to approximate each time-dependent component $v_i(t)$ using an ansatz built from a truncated randomized basis. The most common example, and the one we use below, involves a finite set of Fourier modes with randomly chosen frequencies and phases.
For each control parameter $v_i(t)$, we employ the following ansatz:
\begin{equation}
v_i(t)=A_{i,0} + \sum_{n=1}^{N_{\omega}} \left[ A_{i,n} \cos (\omega_{i,n} t) + B_{i,n} \sin (\omega_{i,n} t) \right].
\end{equation}

We've chosen a distinct set of frequencies $\{\omega_1,...,\omega_{N_{\omega}}\}$ for each component $v_i(t)$, defined as $\omega_{i,n}=n\pi/t_f + \delta \omega_{i,n}$, where $\delta \omega_{i,n}= 2 \pi \delta_{i,n}/t_f$ (with $\delta_{i,n}\in [0,1]$). Randomization is a key ingredient for accelerating convergence. In our CRAB algorithm implementation, these frequencies are fixed during the initialization stage and then used throughout successive iterations.

Within this approach, each coupling $v_i(t)$ is represented by a finite-dimensional vector $(A_{i,0},A_{i,1},...,A_{i,N_{\omega}},B_{i,1},...,B_{i,N_{\omega}})$, as the set of frequencies $\omega_{i,n}$ is treated as fixed variables. The global cost functional $J[U,H]$ from Eq.~\eqref{eq:costfunction} can therefore be viewed as a simpler function $\mathcal{F}(A_{i,0},A_{i,1},...,A_{i,N_{\omega}},B_{i,1},...,B_{i,N_{\omega}})$ of $N_{\rm tot}=(2 N_{\omega}+1) N_C$ variables. Finding the optimal control path then reduces to a simpler problem of multi-variable function minimization. Increasing the number of modes in the randomized basis expands the space of functions explored by the algorithm, leading to a better approximation of the optimal solution, but also comes with significant computational overhead.

We present the results from the CRAB optimization using the cost functional with weights $(p_1=0.5, p_2=0.1)$, which yielded the best outcomes for this routine. The optimization was conducted using a direct-search method to minimize the multi-variable function. This function is derived by solving the Schrödinger equation for the evolution operator, utilizing the set of parameters $(A_{i,0},A_{i,1},...,A_{i,N_{\omega}},B_{i,1},...,B_{i,N_{\omega}})$ that define each coupling's temporal profile.

In our case, we set $N_{\omega}=10$ and $N_C=16$, meaning the minimization was performed with respect to $N_{\rm tot}=336$ variables. As this is a computationally demanding task, we used a smaller number of time bins ($N_t=100$) than in the PMP optimization. Once the pulses were obtained from the optimization routine, we performed a linear interpolation and resolved the Schrödinger equation with $N_t=10^4$ time bins to ensure accurate estimates of the QSL and quantum fidelity.

The optimization results for the four quantum gates, $U_{\rm QFT}$, $U_{\rm ``QFT''}$, $U_{\rm Hadamard}$ and $U_{\rm CNOT}$ are presented in Table \ref{tab:CRABQSLOptimization}. Each optimization consumed a total computing time of approximately 12 hours on a personal laptop, significantly more than the 20 minutes required for the Pontryagin optimization routine.

The trade-off between quantum speed limit and quantum fidelity from the CRAB optimization, as shown in Table \ref{tab:CRABQSLOptimization}, is slightly degraded compared to the Pontryagin optimization (see Table \ref{tab:OptFidelityPlane} and Table \ref{tab:OptPlanePhase}). We attribute this reduced performance to the slower convergence of the CRAB process when handling a large number of couplings. The results depend again on the considered target gate: the CRAB optimization yields an implementation close to the QSL saturation limit for the Hadamard gate, but only delivers a quantum path with a low QSL efficiency for the CNOT gate.

 \begin{table}[htbp]
    \centering
    \begin{tabular}{|l|c|c|c|c|c|}
        \toprule
        \textbf{Gate} & \textbf{Fidelity} &   $J_{\rm Plane}[U,H]$ & $J_{\rm Phase}[U,H]$ & \:  $\overline{\eta}$ \:    \\
       \hline
         $U_{\rm QFT}$     & 0.9993    & 0.054 & 0.045           &  0.951                  \\
        $U_{\rm ``QFT''}$  & 0.9992 & 0.051 & 0.054               & 0.952                   \\
           $U_{\rm Hadamard}$   & 0.9999 & 0.003   &      0.000           &  0.998                     \\ 
           $U_{\rm CNOT}$   &  0.9975       & 0.279 & 0.226           &  0.707              \\
        \hline
    \end{tabular}
\caption{Fidelity optimization with the CRAB routine while minimizing both QSL cost functionals with weights $(p_1=0.5,p_2=0.1)$ in a full control scenario over $16$ couplings in a 4-dimensional system. Optimization realized with a discretization over $N_t=100$ time bins, with $500.000$ iterations for the gates $U_{\rm QFT}$,$U_{\rm ``QFT''}$ and $700.000$ iterations for the gates $U_{\rm Hadamard}$,$U_{\rm CNOT}$.}
\label{tab:CRABQSLOptimization}
\end{table}

\section{CONCLUSION}

To conclude, we have adapted the derivation of the quantum speed limit developed by Vaidmann~\cite{Vaidmann92} for the evolution of single states in the context of quantum unitary operators. This derivation highlights that the quantum operators must satisfy specific geometric conditions. Precisely, the time derivative of the evolution operator must always lie in a plane spanned by the initial operator and the current operator in order to saturate the MT-QSL, and simultaneously fulfill a phase condition. Since the evolution operator must also remain unitary at all times, this imposes stringent constraints, and the MT-QSL cannot always be saturated. More generally, the distance of the evolution operator (or the instantaneous quantum state) from this plane as well as the integrated phase provide a quantitative measure of the deviation from QSL optimality.

Building on these geometric considerations, we've developed an optimal control algorithm designed to account for geometric constraints to best approach the Quantum Speed Limit (QSL). We've tested our approach using two distinct optimization procedures: the Pontryagin Maximum Principle (PMP) and the Chopped Randomized Basis (CRAB) quantum optimization. We investigated various quantum gates relevant to quantum information processing in a two-qubit system.

Our optimization method significantly reduces the departure from QSL optimality while maintaining an excellent level of quantum fidelity. Typically, for the quantum gates we considered, a few parts per thousand in quantum fidelity can be traded for a few percent increase in quantum speed efficiency. Such QSL optimization could lead to significant improvements in quantum devices where rapid implementation of quantum gates is critical.

In quantum spaces of arbitrary dimensions, proving the existence of a quantum trajectory that reaches a given target quantum gate and saturates the QSL can be a challenging problem. By presenting suitable time-dependent quantum protocols that achieve the target gate within a close distance, our optimization procedure provides lower bounds on the optimal average QSL efficiency that can be attained.

Our method could be effectively applied in conjunction with shortcut-to-adiabaticity techniques to improve the QSL efficiency of such quantum protocols~\cite{Sarandy15}.
Given that the existence of a speed limit is fundamentally connected to energy and entropy exchange rates through a time-information uncertainty relation \cite{Nicholson2020}, this methodology extends beyond quantum systems. It can, for instance, include classical stochastic systems governed by Liouvillian dynamics~\cite{QSLQuantClass18,QSLNonQuantum18}.

\emph{Acknowledgements}
 We thank the support from the Institut Universitaire de France, from the ANR project QuCoBEC (ANR-22-CE47-0008-02) and from the CAPES-COFECUB (20232475706P) program. F.I. acknowledges support from the Brazilian agencies CNPq (305638/2023-8), FAPERJ (210.570/2024), and INCT-IQ (465469/2014-0). 

\section*{APPENDIX A: EXPRESSION OF THE COST FUNCTIONALS}

The running plane and phase cost functions $f_{\rm Plane}(U(t),H(t),t)$ and $f_{\rm Phase}(U(t),H(t),t)$ [Equations \eqref{eq:planecondition} and \eqref{eq:phasecondition}] are expressed below in terms of real-valued quantities. Let us define $Z(t)=\langle U(t),\mathbb{I} \rangle \langle \mathbb{I}, U_{\perp}(t) \rangle$.

\begin{widetext}
 \begin{eqnarray}
|| \Delta U_{\perp \perp}(t)||^2  & = & 1- \frac {1} {d^2 \Delta H^2} \left[ \left( {\rm Tr}[H_R U_R- H_I U_I] - \frac {1} {d} {\rm Tr}[H_R] {\rm Tr}[U_R] \right)^2+ \left( {\rm Tr}[H_R U_I+ H_I U_R] - \frac {1} {d} {\rm Tr}[H_R] {\rm Tr}[U_I]\right)^2 \right], \nonumber \\
& \: & - \frac {1} {d^2} \left( {\rm Tr}[U_R]^2+{\rm Tr}[U_I]^2 \right) \nonumber \\
 {\rm Re}[Z(t)]   & = & \frac { 1} { d^2   \Delta H} \left[  {\rm Tr}[U_I] {\rm Tr}[H_R U_I+ H_I U_R]+{\rm Tr}[U_R] {\rm Tr}[H_R U_R- H_I U_I] - \frac 1 d {\rm Tr}[H_R] \left( {\rm Tr}[U_R]^2 +{\rm Tr}[U_I]^2 \right) \right] . \nonumber \\
  {\rm Im}[Z(t)]   & = &  \frac { 1} { d^2   \Delta H} \left[ - {\rm Tr}[U_I] {\rm Tr}[H_R U_R - H_I U_I]+{\rm Tr}[U_R] {\rm Tr}[H_I U_R + H_R U_I] \right]
\end{eqnarray}
\end{widetext}
With $|\langle \mathbb{I}, U(t) \rangle|^2= \frac {1} {d^2} \left( {\rm Tr}[U_R]^2+{\rm Tr}[U_I]^2\right)$, one readily obtains $f_{\rm Plane}(U(t),H(t))$~\eqref{eq:planecondition}. The phase running cost function reads $f_{\rm Phase}(U(t),H(t))= {\rm Re}[Z(t)]^2/ ({\rm Re}[Z(t)]^2+{\rm Im}[Z(t)]^2)$. To compute the  extra driving terms associated to the derivatives of the cost functional, we use that 
$\frac {d {\rm Tr}[A X]} {d X}=A^T$ (where $^T$ is the matrix transpose), together with the properties that $H_R^T=H_R$ and $H_I^T=-H_I$.

\section*{APPENDIX B: DERIVATION OF THE EQUATION FOR THE ADJOINT OPERATOR DYNAMICS}

Let us now put the dynamical equation for the adjoint $\mathbf{p}(t)$ in a more explicit form. From Eqs.(\ref{eq:Pontryagin},\ref{eq:Hamiltoneqs}), one finds:
\begin{equation}
\label{eq:adjoint}
\dot{\mathbf{p}}= \frac {1} {t_f} \frac {\partial f_0(\mathbf{x}(t),\mathbf{v}(t),t)} {\partial \mathbf{x}}-  \frac {\partial} {\partial \mathbf{x}} \left[ \mathbf{p}(t) \cdot  f(\mathbf{x}(t),\mathbf{v}(t),t) \right].
\end{equation}
If one folds the $2N^2$ column vector $\mathbf{p}(t)$ into a $2N \times N$ matrix $\mathbf{P}(t)$ (maintaining the same correspondence between $\mathbf{P}(t)$ and $\mathbf{p}(t)$ as between $\mathbf{X}(t)$ and $\mathbf{x}(t)$), one has:
\begin{eqnarray}
&&\mathbf{p}(t) \cdot    f(\mathbf{x}(t),\mathbf{v}(t),t) =\nonumber \\
&&  \mathbf{P}(t)^{T} \left( \begin{array}{cc} H_I(\mathbf{v}(t)) & H_R(\mathbf{v}(t)) \\
- H_R(\mathbf{v}(t)) &  H_I(\mathbf{v}(t)) \end{array} \right) \mathbf{X}(t) .
\end{eqnarray}
 where $\mathbf{P}(t)^{T}$ is the transpose of $\mathbf{P}(t)$.

Equation \eqref{eq:adjoint} can then be recast as a differential equation for the matrix $\mathbf{P}(t)$:  
\begin{eqnarray}
&& \dot{\mathbf{P}}   =   \frac {1} {t_f}  \left( \begin{array}{c} \frac {\partial f_{0}(\mathbf{U}(t),\mathbf{v}(t))} {\partial U_R} \\ \frac {\partial f_{0}(\mathbf{U}(t),\mathbf{v}(t))} {\partial U_I} \end{array}  \right) \nonumber \\
&&+ \frac {1} {\hbar} \left( \begin{array}{cc} H_I(\mathbf{v}(t)) & H_R(\mathbf{v}(t)) \\
 -H_R(\mathbf{v}(t)) &  H_I(\mathbf{v}(t)) \end{array} \right)  \mathbf{P}(t) .
\end{eqnarray}
We have used the Hermiticity of the Hamiltonian, i.e. $H_{R}(\mathbf{v}(t))^T = H_{R}(\mathbf{v}(t)) $
and $H_{I}(\mathbf{v}(t))^T = -H_{I}(\mathbf{v}(t))$. The second term in the dynamical equation for the adjoint vector $\mathbf{P}(t)$ is identical to the term driving the operator $\mathbf{X}(t)$, and thus corresponds to the Hamiltonian evolution for the adjoint evolution operator $U^P(t)=U_R^P(t)+i U_I^P(t)$ constructed from the adjoint vector (setting $\mathbf{P}(t)=(U_R^P(t),U_I^P(t))^T$). Summing up, one arrives at Eq.~\eqref{eq:adjointpropagationequation}.


\bibliography{QSLoperatorRef}

\begin{thebibliography}{48}
\bibitem [{Foo()}]{FootnoteQSL1}%
  \BibitemOpen
  \href@noop {} {}\bibinfo {note} {Alternatively, one may also generate
  directly the target quantum gate $U_{\rm target}$ without any additional
  phase factor by defining the terminal cost function as $\Phi(U(t_f),t_f)= 1 -
  {\rm Re}[\langle U_{\rm target}, U(t_f) \rangle]^2$ instead of
  $\Phi(U(t_f),t_f)= 1 - |\langle U_{\rm target}, U(t_f)
  \rangle|^2$.}\BibitemShut {Stop}%
\end{thebibliography}

\end{document}